# Using data network metrics, graphics, and topology to explore network characteristics

A. Adhikari[1], L. Denby[1], J. M. Landwehr[1] and J. Meloche[1]

*Avaya Labs*

**Abstract:** Yehuda Vardi introduced the term *network tomography* and was the first to propose and study how statistical inverse methods could be adapted to attack important network problems (Vardi, 1996). More recently, in one of his final papers, Vardi proposed notions of metrics on networks to define and measure distances between a network's links, its paths, and also between different networks (Vardi, 2004). In this paper, we apply Vardi's general approach for network metrics to a real data network by using data obtained from special data network tools and testing procedures presented here. We illustrate how the metrics help explicate interesting features of the traffic characteristics on the network. We also adapt the metrics in order to condition on traffic passing through a portion of the network, such as a router or pair of routers, and show further how this approach helps to discover and explain interesting network characteristics.

## 1. Introduction

Problems in a wide range of fields can be expressed in terms of network structures. Examples range across fields such as social relationships, vehicular traffic, the spread of epidemics, and Internet Protocol (IP) data networks. Data that live on a network go well beyond the classical cases-by-variables paradigm in terms of how the data are represented, the formulation of problems to attack, and the wide range of analytical methods that might be useful in one situation or another.

In a ground breaking and influential paper, Yehuda Vardi [7] identified the problem area of estimating source-destination traffic intensities from link count data and proposed how statistical inverse methods could be adapted to attack this problem, coining the term *network tomography*. More recently, in one of his final papers, Vardi [8] raised notions of defining and using metrics of one type or another in the context of network tomography problems.

Our interest is data networks, in particular test traffic mimicking Voice-over-IP (VoIP) traffic. Data obtained from even limited testing on a corporate network of realistic size are massive and complicated, so some sort of data reduction methods are a natural topic to consider among a range of analytical options when attacking certain problems. Measures of distance and similarity provide one way to summarize and reduce the raw data. In this paper, we explore how adapting some of Vardi's ideas on metrics can supply useful insights for interesting problems relating to VoIP traffic flows. We think of the methods as data analysis and data summarization tools, possibly useful in initial analytical steps, apart from whether or not more extensive analytical capabilities using the mathematical properties of metrics are taken advantage of or not.

[1]Avaya Labs, 233 Mt Airy Road, Basking Ridge, New Jersey 07920, USA, e-mail: akshay@avaya.com; ld@avaya.com; jml@avaya.com; jmeloche@avaya.com

It is natural to think of a data network in terms of a graph in which the vertices represent either routers in the core of the network or communication endpoints at the periphery of the network, and the edges in the graph represent network links on which traffic can flow. A whole range of distance measures can be thought of in this context. For example, the distance from one vertex to another can be defined to be the length of the shortest path between the two vertices; that is, the distance is the minimum number of hops required to traverse through edges from one vertex to the other in the graph. If the edges have a direction, this is a directed graph and this distance measure is not necessarily symmetric. If the edges are all bi-directional, however, this distance is symmetric.

The above idea of measuring the distance between vertices by the length of the shortest path can clearly be extended to take edge performance values such as latency, loss or jitter into account. Indeed, a distance measure from $x$ to $y$ can be defined by minimizing the sum of edge values over all paths that go from $x$ to $y$. The original measure corresponds to the case where all the values are 1.

Real data networks have the additional structure of routing. Routing is the process by which messages sent by one communication endpoint find their way through the network to the destination endpoint. Routing is performed by devices called routers and is partly dictated by protocols such as OSPF. These protocols, in turn, are based on metrics such as the above shortest path and also other considerations such as security or policy.

One way to describe routing is to specify the paths through the network between the endpoints of interest. These paths can be thought of as sequences of vertices, starting with the source endpoint and ending with the destination endpoint. Routing in real networks is often asymmetric (the path from B to A is not simply the reversed sequence of vertices in the path from A to B), it is time dependent, and it depends on the nature of the end-to-end traffic (e.g., www traffic is not necessarily routed along the same path as VoIP traffic).

Given this additional routing structure, there is a need to go beyond approaches that assign distances between vertices. Vertices are routers and endpoints, and the edges between vertices are the basic elements of interaction in a network. In principle, a path through the network could be characterized either by a sequence of routers and/or by a sequence of edges. From an information standpoint, however, the sequence of edges constitutes a more complete description of the path since it amounts to a sequence of router interfaces rather than simply a sequence of routers. Moreover, including routers alone seems naive in that having interaction or not between adjacent routers is an important network characteristic – that is, are the routers connected by a network link or edge? For these reasons, we prefer to think of paths as sequences of edges rather than as a sequence of vertices. These points motivate considering topics around distances between the edges of a network.

Clearly many notions of distance could be created and studied. Our interest is around the similarity of traffic across various links, or across certain sets of links (paths) going from one communication endpoint to another endpoint. Thus, we are primarily interested in distances (or similarities) between pairs of edges, rather than between pairs of vertices (nodes).

Taking this perspective, we investigate ways to construct and use distances that are defined between pairs of edges in the network. The definition of the distances is conditional on the following network traffic characteristics: the quantity and intensity of the end-to-end communication traffic sent through the network; the paths taken by this traffic; and the dependence of the routing over time. The values of the metric could include performance measures of the end-to-end traffic.

Section 2 presents the definitions needed for our approach. In Section 3, we provide some initial examples and graphical displays to visualize the set of distances between pairs of edges in the network. Section 4 proposes an application of the distances-between-edges calculations in which the communication endpoints are automatically assigned in a hierarchy that reflects end-to-end performance. In Section 5, we describe a second application of the distances in which the entire network is automatically classified into one of several canonical types based on the distribution of the distances between the edges. Section 6 provides a few concluding comments.

There are, of course, many statistical problems related to IP network data. For discussions of other problems, an issue of *Statistical Science* contains three companion review articles that survey many recent developments in the field: see Duffield [5], Castro et al. [4], and Cao et al. [3], each of which also points to many further references. Our research interests have grown out of the needs to develop technology to test and assess enterprise networks for VoIP Quality of Service (QoS); see Bearden, et al. [1, 2] and Karacali, et al. [6].

## 2. Definitions

Consider a data network including a set of communication endpoints used as traffic sources and destinations, along with a specific pattern of test traffic generated among pairs of these devices. Conditional on this setup, we wish to calculate distances among the edges in the network based on the similarities of the traffic flowing across the edges, in ways that we will make precise.

Traffic is routed according to a given routing matrix $A$. The rows of $A$ represent the edges in the network and the columns represent the paths of interest. The paths correspond to each source-destination pair of endpoints that is tested. The element at row $e$ and column $p$ is

$$(2.1) \qquad a_{ep} = \begin{cases} w_p \text{ if } e \in p \\ 0 \text{ if } otherwise \end{cases}$$

where $w_p$ is whatever path level quantity is of interest. Examples of $w_p$ could be the indicator variable 1, or the path frequency within the test plan, or an end-to-end performance measure such as delay or loss on this path. On the other hand, $w$ could be defined to be edge effects derived from end-to-end values such as those resulting from network tomography calculations on delays.

Vardi proposed to measure the distance between edges in terms of the rows of the matrix $A$. Let $A_e$ be the row of $A$ that corresponds to the edge $e$. Specifically, he proposed

$$(2.2) \qquad d(e, f) = |A_e - A_f|$$

as a distance between edges when $w_p = 1$ and there is exactly one path for each source-destination pair and he demonstrated that $d$ satisfies the properties of a metric.

We propose to explore variations on this original idea. For edges $e$ and $f$, define

$$(2.3) \qquad d_1(e, f) = |A_e - A_f| / \max(|A_e|, |A_f|)$$

in which the difference between the two row vectors is normalized by the larger of their norms. Continuing the extension, let $P_e$ be the set of paths containing edge $e$. Let $R \triangle S$ denote the symmetric difference of the two sets $R$ and $S$, so

$$(2.4) \qquad R \triangle S = (R \cup S) - (R \cap S)$$

We also define

$$(2.5) \qquad d_2(e,f) = \frac{\sum_{p \in P_e \triangle P_f} w_p}{\sum_{p \in P_e \cup P_f} w_p}$$

which reduces to

$$(2.6) \qquad d_3(e,f) = \frac{|P_e \triangle P_f|}{|P_e \cup P_f|}$$

when $w_p \equiv 1$. Vardi's $d$ is the numerator of $d_3$ when there is exactly one path for each source-destination pair. We find that normalizing his quantity by the denominator of $d_3$, a quantity essentially representing the total frequency with which edges $e$ and $f$ appear in the paths under test from this traffic configuration, gives an intuitively more plausible measure of distance. Moreover, in $d_2$ we extend the concept that Vardi proposed to the case where multiple paths from source to destination are observed, a common situation in real networks at least over a sufficient period of time. This extension also opens up the possibility of incorporating alternative path level quantities into the distance measure.

## 3. Examples

In this section we give two examples to better understand the edge distance measure. We focus on the similarity measure $s = 1 - d_2$. We start with a large example, then specialize it to a subgraph in the smaller second example to illustrate the calculations. The data come from a real, world-wide corporate network. It consists of paths between 28 endpoints deployed on the corporate network.

The paths are obtained with the common traceroute utility which was run between randomly selected pairs of endpoints about 1.3 million times in a period of 10 days which amounts to not quite 2 tests running every second. The traceroute test works by sending packets from source to destination with increasing time to live (TTL) values. The time to live value of a packet is the number of routers the packet is allowed to traverse on its way to the destination. Each intervening router decrements the time to live value and when it has reached the value 0, the router sends its address back to the source in a TTL Exceeded (ICMP) message. The result is a sequence of addresses from source to destination at increasing distances from the source. The testing process resulted in 1825 paths observed at one time or another.

Figure 1 is a histogram of the path frequencies. If routing never changed, we would have observed only $756 = 28 \times 27$ of them, each of which would have occurred about 1900 times.

Figure 1 reveals an interesting aspect of the paths on this network. There are accumulations of frequencies around 1800, and at binary divisions of 1800, with the bumps at 900 and 450 clearly visible. This suggests that there are certain paths that occur each time a certain source-destination pair is tested, while for other S-D pairs there are two or four paths that occur roughly equally frequently.

**Path frequency**

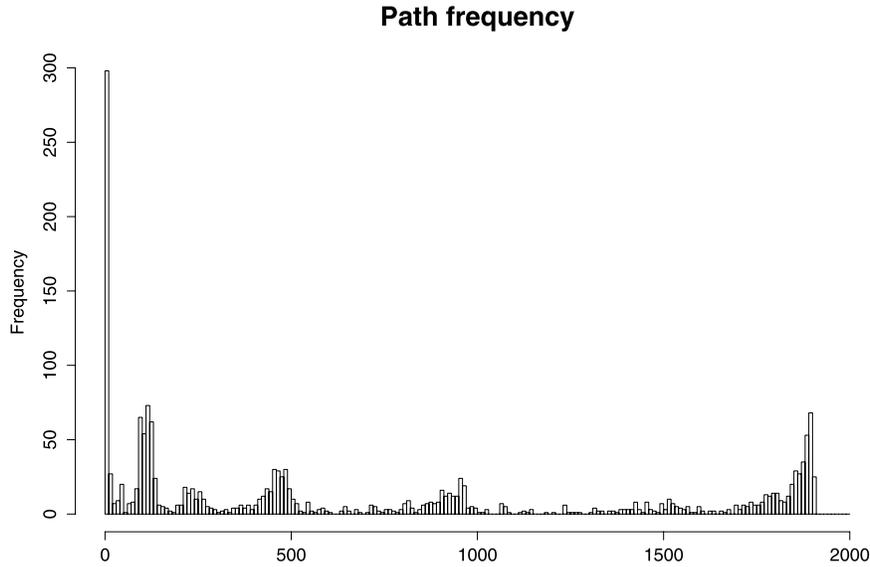

FIG 1. *Distribution of path frequency throughout collection period.*

Table 1 shows the number of observed paths for each of the source-destination pairs. Note that this table shows only 27 rows and columns since no traceroute data originated from one of the endpoints so that row and column were eliminated. The first two endpoints have a large number of paths to destination due to load balancing and fail-over nodes near their location. Load balancing nodes are used in turn according to some schedule such as round robin in order to equalize traffic over alternative network links. Fail-over nodes are only used when a primary node fails. Both types of nodes are the primary explanations for the multiplicity of the paths from source to destination. One remarkable aspect of this multiplicity is that it is not symmetric at all.

The second example involves only two endpoints, both in New Jersey with one in Lincroft and the other in Basking Ridge. For this example, the routing matrix $A$ is formed by restricting the data to these two endpoints. Figure 2 shows this restricted network. The endpoints in Lincroft and Basking Ridge are shown as triangles at the left and right edge of the network and the circles represent the routers involved in the end-to-end traffic between the two endpoints. The direction of the traffic is shown as an arrow on the edge or as a dot for bidirectional edges.

Figure 2 shows the network topology but fails to express how this topology is used in the end-to-end communication between the two endpoints. The corresponding routing information is reported in Table 1 which amounts to the weighted routing matrix $A$. In this case, we have three paths, two from Lincroft to Basking Ridge and one in the reverse direction. Table 1 reports how frequently each edge is used in each of the paths during the 10 day period. Over time, the network kept shifting between Path 1 and Path 2 for tests from Lincroft to Basking Ridge, while tests in the opposite direction always followed Path 3. The entries in the table are the $w_1, w_2,$ and $w_3$ which we plug into Equation 2.5 to calculate the edge distances.

The edge traffic similarity $1 - d_2$ can be used to measure the similarity of the end-to-end traffic that goes through a pair of edges. By way of example,

(3.1) $$s(\text{Lin} - \text{B}, \text{I} - \text{G}) = 0$$

TABLE 1
*Number of paths per source-destination pair*

|    | 0 | 1 | 2  | 3 | 4  | 5 | 6  | 7  | 8 | 9  | 10 | 11 | 12 | 13 | 14 | 15 | 16 | 17 | 18 | 19 | 20 | 21 | 22 | 23 | 24 | 25 | 27 |
|----|---|---|----|---|----|---|----|----|---|----|----|----|----|----|----|----|----|----|----|----|----|----|----|----|----|----|----|
| 0  |   | 1 | 12 | 6 | 11 | 6 | 10 | 10 | 6 | 10 | 1  | 10 | 10 | 10 | 10 | 10 | 10 | 10 | 10 | 10 | 10 | 10 | 6  | 6  | 6  | 1  | 1  |
| 1  | 1 |   | 12 | 6 | 10 | 6 | 10 | 10 | 6 | 10 | 1  | 10 | 10 | 10 | 10 | 10 | 10 | 10 | 10 | 10 | 10 | 10 | 6  | 6  | 6  | 1  | 1  |
| 2  | 2 | 4 |    | 4 | 2  | 4 | 2  | 2  | 2 | 2  | 2  | 2  | 2  | 2  | 2  | 1  | 2  | 2  | 2  | 2  | 2  | 2  | 1  | 2  | 10 | 2  | 2  |
| 3  | 2 | 2 | 4  |   | 1  | 3 | 2  | 2  | 3 | 2  | 3  | 3  | 2  | 2  | 3  | 3  | 2  | 2  | 2  | 1  | 1  | 5  | 2  | 3  | 1  | 1  | 1  |
| 4  | 8 | 8 | 8  | 4 |    | 4 | 4  | 4  | 4 | 8  | 4  | 1  | 4  | 4  | 4  | 4  | 1  | 4  | 4  | 4  | 4  | 4  | 1  | 1  | 4  | 4  | 1  | 1 |
| 5  | 4 | 4 | 4  | 2 | 2  |   | 2  | 2  | 2 | 2  | 1  | 2  | 2  | 2  | 2  | 1  | 2  | 2  | 2  | 2  | 2  | 2  | 2  | 2  | 4  | 1  | 1  |
| 6  | 2 | 2 | 4  | 4 | 2  | 4 |    | 3  | 2 | 2  | 2  | 2  | 2  | 2  | 2  | 2  | 2  | 2  | 1  | 2  | 1  | 2  | 2  | 4  | 2  | 2  | 2  |
| 7  | 3 | 3 | 3  | 2 | 2  | 2 | 2  |    | 1 | 1  | 1  | 1  | 1  | 1  | 1  | 2  | 2  | 2  | 2  | 2  | 2  | 3  | 1  | 2  | 1  | 1  | 2  |
| 8  | 2 | 1 | 2  | 1 | 2  | 1 | 1  | 1  |   | 1  | 1  | 2  | 1  | 2  | 2  | 1  | 1  | 1  | 1  | 1  | 1  | 2  | 1  | 2  | 5  | 1  | 2  |
| 9  | 1 | 1 | 2  | 1 | 1  | 1 | 1  | 2  | 1 |    | 1  | 1  | 1  | 1  | 1  | 1  | 1  | 1  | 1  | 1  | 1  | 1  | 1  | 1  | 4  | 1  | 1  |
| 10 | 1 | 2 | 4  | 2 | 1  | 2 | 1  | 2  | 1 | 2  |    | 1  | 1  | 1  | 1  | 1  | 1  | 1  | 1  | 1  | 1  | 1  | 2  | 3  | 4  | 2  | 1  |
| 11 | 2 | 2 | 2  | 1 | 1  | 1 | 1  | 1  | 2 | 1  | 1  |    | 1  | 1  | 1  | 1  | 1  | 1  | 1  | 1  | 1  | 1  | 1  | 2  | 4  | 1  | 1  |
| 12 | 1 | 1 | 2  | 2 | 1  | 2 | 1  | 1  | 1 | 1  | 1  | 1  |    | 1  | 1  | 1  | 1  | 1  | 1  | 1  | 1  | 1  | 1  | 2  | 4  | 1  | 1  |
| 13 | 1 | 2 | 2  | 1 | 1  | 1 | 1  | 1  | 2 | 1  | 1  | 1  | 1  |    | 1  | 1  | 1  | 1  | 1  | 1  | 1  | 1  | 1  | 1  | 4  | 1  | 1  |
| 14 | 2 | 1 | 2  | 1 | 1  | 2 | 1  | 1  | 1 | 1  | 1  | 2  | 1  | 1  |    | 1  | 1  | 1  | 1  | 1  | 1  | 1  | 1  | 1  | 4  | 1  | 1  |
| 15 | 2 | 1 | 2  | 1 | 1  | 2 | 1  | 2  | 2 | 2  | 2  | 1  | 1  | 1  | 1  |    | 1  | 1  | 2  | 1  | 1  | 1  | 1  | 1  | 4  | 1  | 1  |
| 16 | 2 | 2 | 2  | 2 | 1  | 2 | 1  | 2  | 1 | 1  | 1  | 1  | 1  | 1  | 1  | 1  |    | 1  | 1  | 1  | 1  | 2  | 1  | 2  | 4  | 1  | 1  |
| 17 | 2 | 2 | 2  | 1 | 1  | 2 | 1  | 2  | 1 | 1  | 1  | 1  | 1  | 1  | 1  | 1  | 1  |    | 1  | 1  | 1  | 1  | 1  | 2  | 4  | 1  | 1  |
| 18 | 1 | 2 | 2  | 2 | 1  | 2 | 1  | 2  | 1 | 1  | 1  | 1  | 1  | 1  | 1  | 1  | 1  | 1  |    | 1  | 1  | 2  | 1  | 1  | 8  | 1  | 1  |
| 19 | 2 | 1 | 2  | 1 | 1  | 2 | 1  | 2  | 1 | 1  | 1  | 1  | 1  | 1  | 1  | 1  | 1  | 1  | 1  |    | 1  | 2  | 1  | 2  | 8  | 1  | 1  |
| 20 | 2 | 2 | 2  | 2 | 1  | 2 | 1  | 2  | 1 | 1  | 1  | 1  | 1  | 1  | 1  | 1  | 1  | 1  | 1  | 1  |    | 1  | 1  | 2  | 4  | 1  | 1  |
| 21 | 1 | 1 | 4  | 3 | 1  | 3 | 1  | 2  | 2 | 1  | 1  | 1  | 1  | 1  | 1  | 2  | 1  | 2  | 2  | 1  |    | 2  | 3  | 4  | 2  | 1  |    |
| 22 | 1 | 1 | 2  | 2 | 1  | 2 | 1  | 1  | 2 | 2  | 1  | 1  | 1  | 1  | 2  | 1  | 1  | 1  | 1  | 1  | 2  |    | 2  | 1  | 1  | 1  |    |
| 23 | 1 | 1 | 2  | 2 | 2  | 2 | 1  | 1  | 1 | 1  | 1  | 1  | 1  | 1  | 1  | 1  | 1  | 1  | 1  | 1  | 1  | 2  | 1  |    | 8  | 1  | 1  |
| 24 | 1 | 1 | 2  | 1 | 1  | 2 | 1  | 2  | 2 | 1  | 1  | 2  | 1  | 1  | 1  | 1  | 1  | 1  | 1  | 1  | 1  | 1  | 1  | 2  |    | 1  | 1  |
| 25 | 2 | 2 | 2  | 1 | 1  | 1 | 1  | 1  | 1 | 1  | 1  | 1  | 1  | 1  | 1  | 1  | 1  | 1  | 1  | 1  | 1  | 2  | 1  | 1  | 1  |    | 1  |
| 26 | 1 | 1 | 2  | 2 | 1  | 1 | 1  | 2  | 2 | 1  | 1  | 1  | 1  | 1  | 1  | 1  | 1  | 1  | 1  | 1  | 1  | 1  | 1  | 2  | 4  | 1  |    |

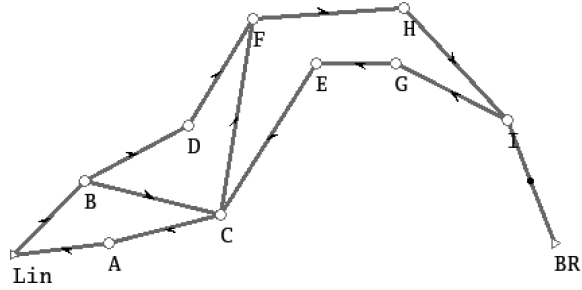

Fig 2. *New Jersey Topology.*

since none of the end-to-end traffic that goes through Lin-B goes through I-G. On the other hand,

$$(3.2) \qquad s(\text{Lin} - \text{B}, \text{B} - \text{D}) = 1336/(0 + 1336 + 552) = 0.71$$

The rest of the similarity values are reported in Table 1. Figure 3 shows the same information graphically. For better visibility, the edges that have similarity of zero to Lin-B are colored grey and the others are shaded in a gradation from pink to red.

We end this section will a full map of the corporate topology which expresses the similarity of traffic between all edges to the Lin-B edge. Figure 4 illustrates the important idea that because traffic is end-to-end, similarities can be found between edges that are far apart in the usual sense of hop distance or geography. Figure 4 shows that overall few edges have some similarity to Lin-B and that those that do have some similarity are widely dispersed in a geographical and in a network sense. Furthermore, manipulating the color rendering on a computer screen (which we cannot convey in a static picture) reveals that among edges that have some similarity to Lin-B, a number of the inter-continental edges of the network have stronger similarity to Lin-B than the intra-continental ones. This is reflecting that there are relatively few cross-continental edges which are left to concentrate the traffic from one continent to another.

Table 1
*New Jersey weighted routing matrix*

| Edge  | Path 1 | Path 2 | Path 3 | Similarity to Lin-B |
|-------|--------|--------|--------|---------------------|
| Lin-B | 552    | 1336   |        | 1.0                 |
| B-C   | 552    |        |        | 0.29                |
| C-F   | 552    |        |        | 0.29                |
| F-H   | 552    | 1336   |        | 1.0                 |
| H-I   | 552    | 1336   |        | 1.0                 |
| I-BR  | 552    | 1336   |        | 1.0                 |
| B-D   |        | 1336   |        | 0.71                |
| D-F   |        | 1336   |        | 0.71                |
| BR-I  |        |        | 1889   | 0                   |
| I-G   |        |        | 1889   | 0                   |
| G-E   |        |        | 1889   | 0                   |
| E-C   |        |        | 1889   | 0                   |
| C-A   |        |        | 1889   | 0                   |
| A-Lin |        |        | 1889   | 0                   |

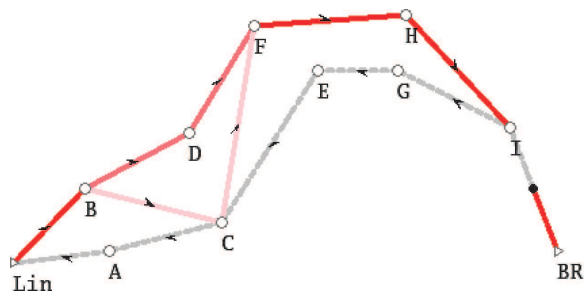

Fig 3. *Edge similarity to Lin-B.*

## 4. Automatic hierarchy of endpoints

Active monitoring of a network involves sending probes to and from each endpoint. Such a pairwise testing of all pairs does not scale easily with increasing number of endpoints since adding an additional endpoint given that there are already $n$ endpoints being used results in adding $2n$ additional pairs to be tested. Thus the number of pairs to be tested grows like $n^2$.

In designing the monitoring plan one must balance network coverage with the intensity of the monitoring data over the links. One does not want to swamp the links, yet one does not want to visit a particular region of the network (or pair) too infrequently. One way to handle this balance is by assigning the endpoints to a hierarchical tree that defines the test scheduling between endpoints. For example, take the network in Figure 4. One possible tree is shown in Figure 5. This tree was specified by the network engineer who is familiar with the routing pattern of the network along with the existence of certain applications on the network, e.g., VoIP or video conferencing.

This hierarchy specifies the scheduling of the tests. Tests are performed consecutively for all pairs in one level of the hierarchy and then between pairs in the next level of the hierarchy. Thus, on the first round there will be tests performed between pairs of EMEA, APAC, US and Backbone regions where one of the endpoints in the leaves below that region is chosen at random to represent the region. After those six tests are completed then the 10 tests within EMEA are performed, followed by the 10 tests within APAC, followed by an East-West US test, and then the test within the Backbone. Then we go to the next level and perform the 10 tests within the west and the North-South test and lastly the tests within North and within South. This completes one round of testing and we can start over at the top level again. So, this hierarchy defines a round of testing comprised of 64 tests instead of the 378 that it would take to cover all $n(n-1)/2$ pairs of endpoints.

The advantage of this is less traffic on the network and more frequently visited regions of interest.

Setting up this hierarchy by hand requires considerable knowledge of the network and can prove to be time consuming for large networks. Thus, there is a need for an automatic way of determining the hierarchy and assigning the endpoints to each node of the hierarchy. A way to automatically determine the hierarchy is to use the routing incidence matrix (probably collected at a moderate pace between all pairs of endpoints) along with clustering to determine how groups of endpoints combine to form the hierarchy. For example, suppose we take the routing matrix which contains counts of the number of times each link is touch by a path such as in Table 2 and collapse the columns related to the paths between the same source-destination pair of endpoints by summing them and dividing by the total number of times that source/destination pair was used in testing. For example, columns 1 and 2 of Table 2 are replace by a single column with values (1, 0.29, 0.29, 1, 1, 1, 0.71, 0.71, 0, 0, 0, 0, 0, 0). In this case a link entry for a source-destination pair would represent the proportion of times that that link has been touched in the paths between this source-destination pair. The distance between source and destination would be the sum of the resulting column. Thus, in this case the distance is essentially the number of hops between source and destination adjusted for the multiple paths being varying lengths. To symmetrize the distance matrix we use $(\mathbf{D} + \mathbf{D}^T)/2$. It is this symmetric distance matrix that is then used in a straightforward clustering algorithm to determine the hierarchy.

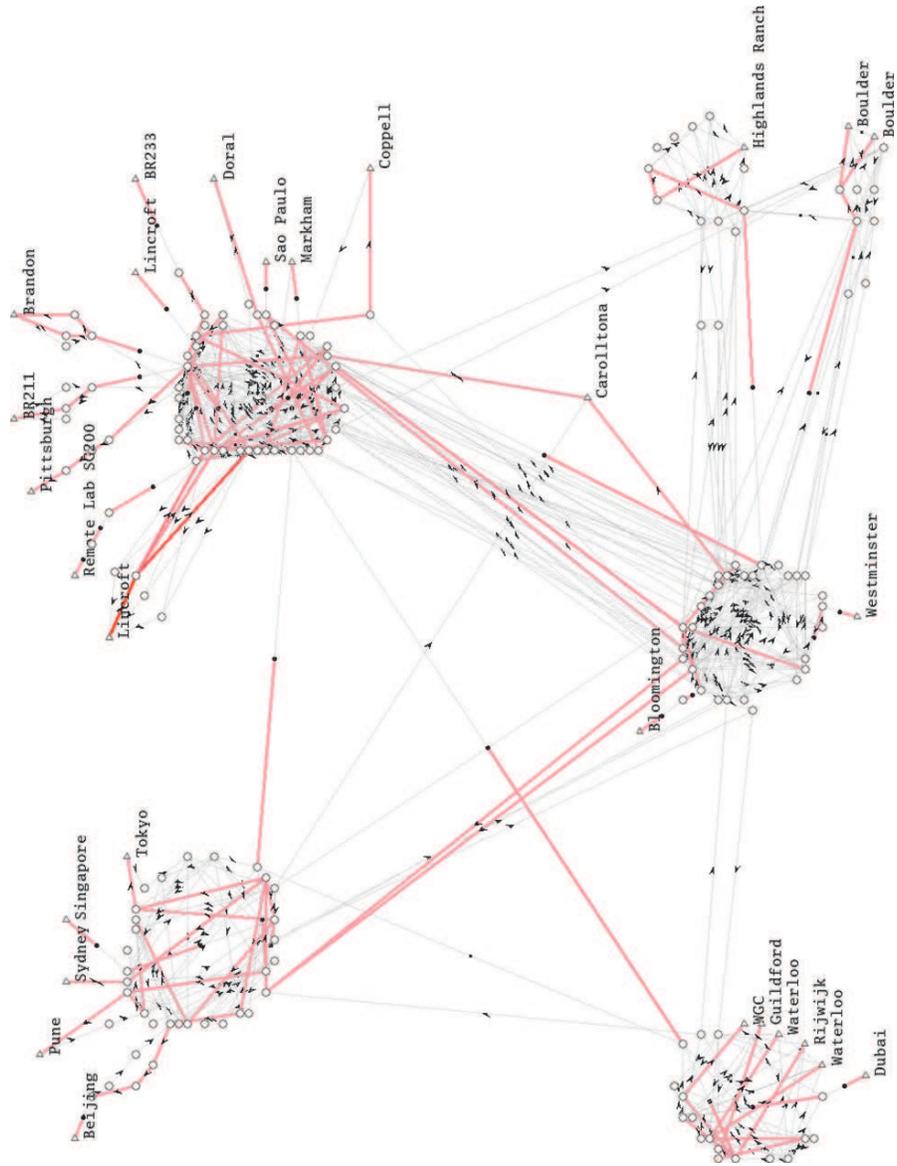

Fig 4. *Edge similarity to Lin-B.*

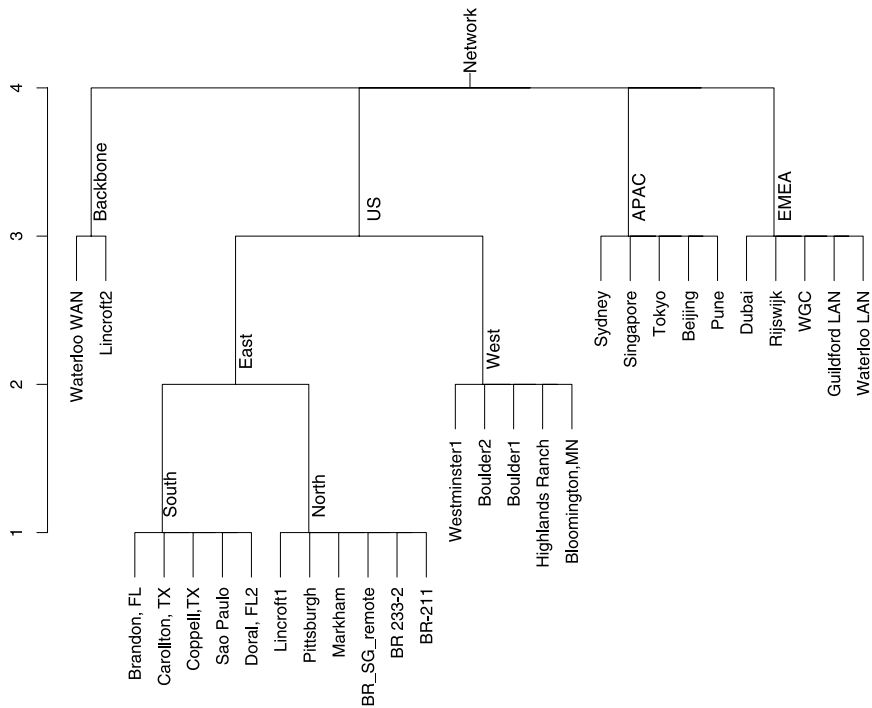

Fig 5. *Manual hierarchy.*

Figure 6 shows the hierarchical clustering based on this simple algorithm. Generally the clustering follows what might be set intuitively by the network administration, such as in Figure 6, thus showing that using this routing matrix as the basis of an automatic hierarchy algorithm is promising. In fact, one might think of enhancing the **A** matrix to reflect the network metrics collected along a path or other network parameters such as throughput or capacity, thus automatically constructing a distance that is not only based on the number of hops but also measures of the performance between the endpoints.

## 5. Network classification

A second application of edge distance is to classify networks according to the distances between the edges in that network. The idea is that this distribution could

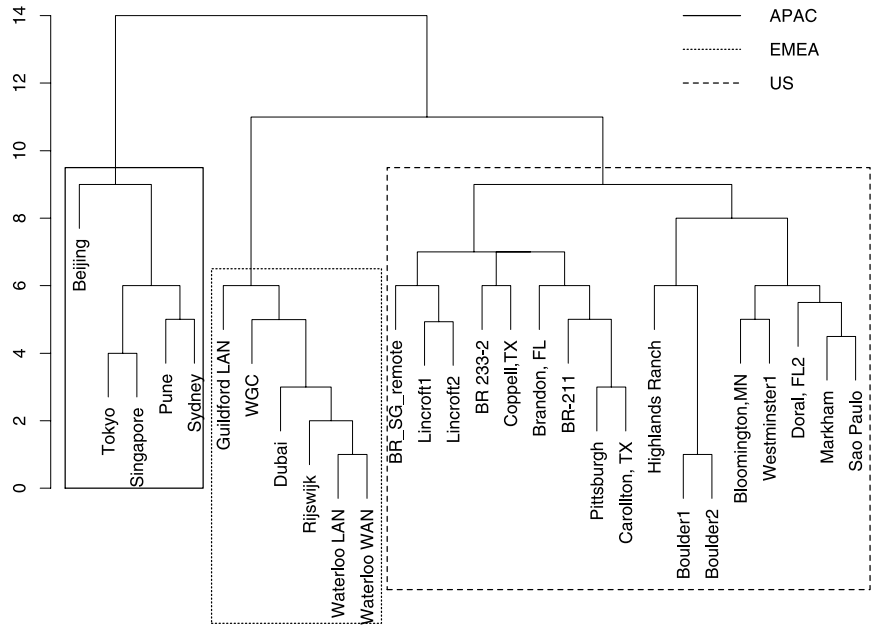

Fig 6. *Automatic hierarchy.*

serve as a signature of the paths on the network. To illustrate how such a signature may be useful, we start by producing the signature distribution for four networks on which there are four fundamentally different routing patterns. The networks are

- ring (unidirectional), where the nodes are disposed in a ring and the links are unidirectional;
- ring (bidirectional), where the nodes are disposed in a ring and the links are bidirectional;
- star, where the nodes are disposed in a star around a middle router;
- mesh, where all nodes connect directly to one another.

Figure 7 illustrates the four networks based on 10 nodes. For each of the four networks, we calculated $d_3$ for all edge pairs and formed a histogram of the resulting similarities. The calculations for $d_3$, however, were made for networks with 100 nodes. Figure 8 displays the resulting histograms for the corresponding similarities. The star and mesh histograms look the same, but while mesh actually has all of its mass at 0, star has a bit of it off 0.

The histograms show as expected that the edges on the star and mesh have (little or) no similarity because distinct edges (hardly) ever carry the same end-to-end traffic. The rings, however, are different and the unidirectional ring, in particular has a lot of similar edges. In fact, similarity for the rings is a function of the hop distance between the edges.

Figure 9 displays the histogram signature for a worldwide corporate network. The signature is similar to the star network, with much but not all of its mass near 0, which tells us that the edges on the corporate network have little or no similarity.

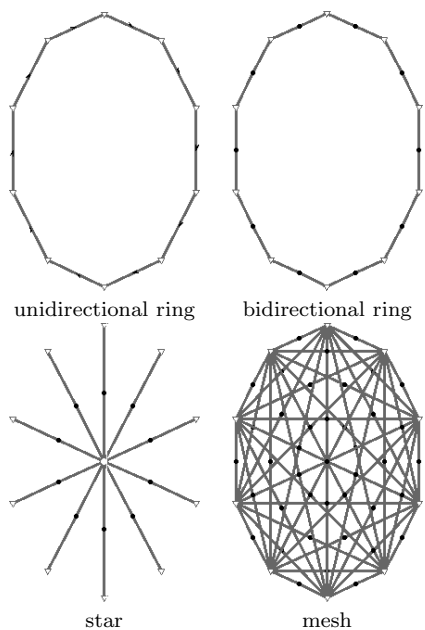

Fig 7. *Candidate networks.*

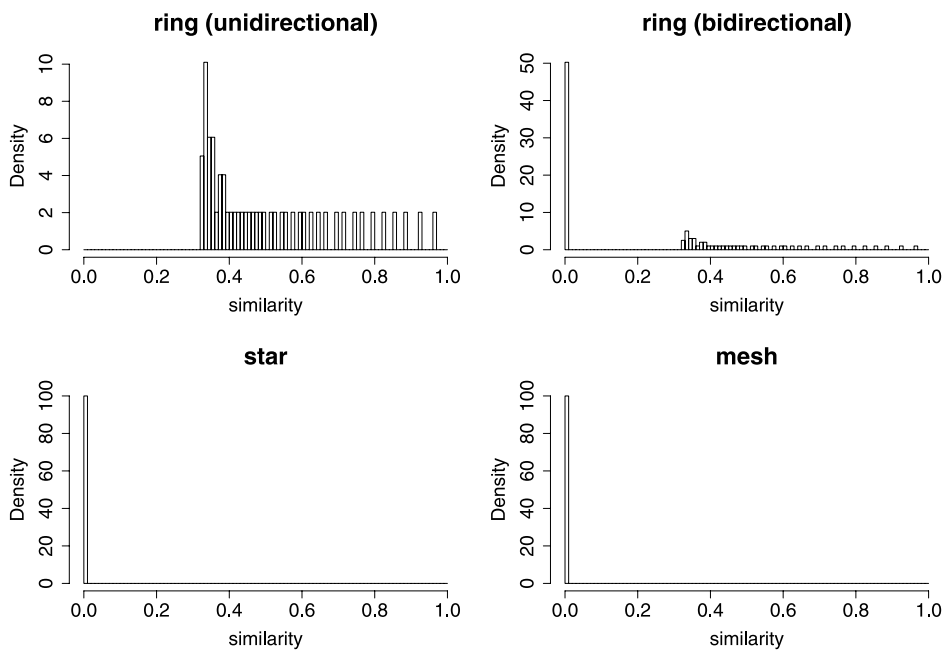

Fig 8. *Distributions of edge similarities.*

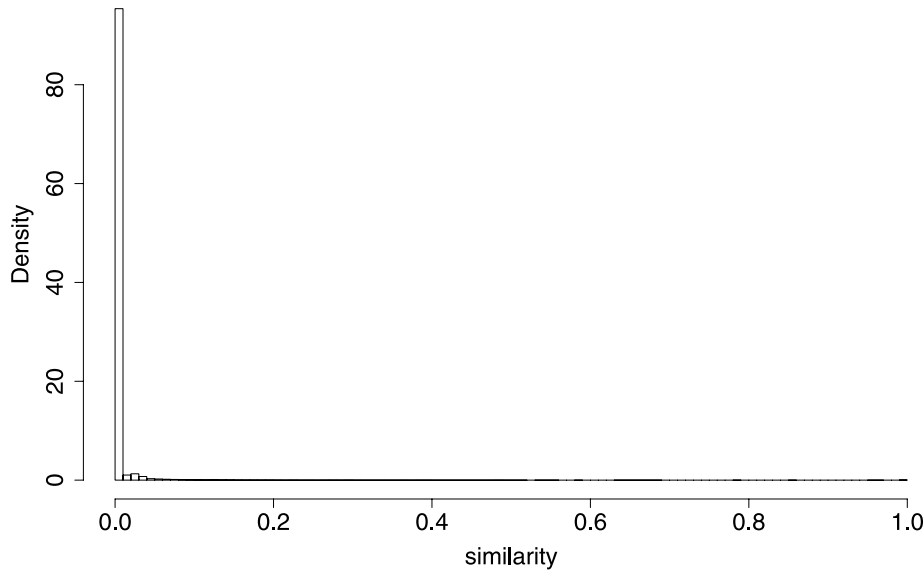

Fig 9. *Distributions of edge distances on the corporate network.*

## 6. Summary and conclusions

Data networks incorporate many features: hops, traffic flow, delay, jitter, loss, topology, random paths, etc. They are complex and the data available to characterize them is voluminous. In order to help study their characteristics and to compare two data networks, Vardi suggested an approach for creating network metrics based on traffic flow conditional on source/destination endpoints.

In this paper we show that the distribution of edge similarities is related to the structure of the network topology. In addition, displaying network metrics on top of the topology is useful since it helps in the following: 1) understanding of traffic flow on the network; 2) identifying multiple paths used in the network; 3) recognizing where topologically separate parts of the network have similar traffic; and 4) recognizing where adjacent links have dissimilar traffic flows. Lastly, using end-to-end traffic flow metrics can lead to automatically creating useful hierarchies among the endpoints that can be used in designing scalable monitoring strategies.

In conclusion, Vardi's idea of using network metrics of traffic flow has merit and is a fruitful topic for current and future research. We thank Yehuda for pointing out the value of using traffic flow metrics to help understand and compare networks.